\documentclass[runningheads]{llncs}
\usepackage{graphicx}
\usepackage{float}
\usepackage{mathptmx} 
\usepackage{amsmath} 
\usepackage{amssymb}  
\usepackage{color}
\usepackage{adjustbox}
\usepackage{multirow}
\usepackage{times}
\usepackage{array}
\usepackage{subfig}
\begin{document}
\title{Discriminative Localized Sparse Representations for Breast Cancer Screening}
%
%
\author{Sokratis Makrogiannis\orcidID{0000-0003-0316-3529} \and Chelsea E. Harris\orcidID{0000-0002-3239-3983} \and Keni Zheng\orcidID{0000-0001-9318-1362}}
%
%
\institute{Division of Physics, Engineering, Mathematics and Computer Science\\ Delaware State University, Dover, DE 19901, USA\\
\email{smakrogiannis@desu.edu}}
\maketitle              
\begin{abstract}
Breast cancer is the most common cancer among women both in developed and developing countries. Early detection and diagnosis of breast cancer may reduce its mortality and improve the quality of life. Computer-aided detection (CADx) and computer-aided diagnosis (CAD) techniques have shown promise for reducing the burden of human expert reading and improve the accuracy and reproducibility of results.
Sparse analysis techniques have produced relevant results for representing and recognizing imaging patterns. In this work we propose a method for Label Consistent Spatially Localized Ensemble Sparse Analysis (LC-SLESA). In this work we apply dictionary learning to our block based sparse analysis method to classify breast lesions as benign or malignant. The performance of our method in conjunction with LC-KSVD dictionary learning is evaluated using 10-, 20-, and 30-fold cross validation on the MIAS dataset. Our results indicate that the proposed sparse analyses may be a useful component for breast cancer screening applications.
\end{abstract}

\section{Introduction}
In this paper we introduce a method for classification of breast masses into benign and malignant states. Breast cancer diagnosis is a significant area of research \cite{Ferlay2010}, and because of its significance, automated detection and diagnosis of breast cancer is a popular field of research \cite{Oliver2010,Verma2010,Pereira2014,Huynh2016,Nagarajan2017}. Early diagnosis has been shown to reduce mortality and significantly improve quality of life. To achieve this, mammograms are used to help detect breast cancer at an early stage. Detection and diagnosis of breast cancer requires high levels of expertise and experience, and is carried out by trained radiologists. Computer-aided diagnosis would reduce the time for diagnosis and improve its reproducibility. 

A widely adopted method for diagnosis and early prediction of breast cancer is the X-ray mammographic test \cite{Misra2010screening}. Hence the research field of computer-aided detection and diagnosis for breast cancer has attracted significant interest. Conventional classification models such as those introduced in \cite{Beura2015,rouhi2015benign,rabidas2017neighborhood,Singh2016,Narvaez2017,George2019} use specific rules to craft features. Texture, shape and intensity features were extracted in \cite{rouhi2015benign}. Among the extracted features, genetic algorithm (GA) selected the most appropriate features. Zernike moments have also been used to extract features due to their ability to well describe the shape of objects \cite{Sharma2019}. In more recent years, state of the art techniques using neural networks have been used to extract and select features \cite{litjens2017survey}. Among these techniques, Convolutional Neural Nets (CNNs) are a popular choice. Key advances in design and application of CNNs \cite{krizhevsky2012imagenet,Szegedy2015} significantly improved the state of the art in object recognition for the imagenet dataset. A commonly used training strategy for neural networks for medical imaging classification applications is transfer learning \cite{litjens2017survey,hepsaug2017using,Zhao2015a}. In \cite{Chougrad2018}, for example, pretrained VGG16, ResNet50, and Inception v3 networks were customized to be applied on different datasets. 

This research focuses on the diagnosis (CADx) of breast cancer lesions as benign or malignant using sparse representation and dictionary learning. Sparse representation has found wide applicability in signal/image processing, computer vision and pattern recognition. Sparse representation methods seek to represent signals using sparse linear approximations of patterns, or atoms that compose the dictionary matrix. These sparse approximations can be used for compression and denoising of signals/images, classification, object recognition, and other applications. A central theme in such techniques is dictionary learning, which studies methods for learning dictionaries that lead to optimal representations according to the application objective. These methods have produced impressive results in various signal and image processing tasks \cite{aharon2006,tosic2011dictionary,jiang2013label,yang2014sparse,zhou2014classification,zhang2015survey,zheng2017,rey2020variations}.
In the recent years, convolutional sparse coding and its relationship with deep learning methods has been widely studied \cite{zhang2015survey,Chang2017,rey2020variations}. 

Despite the interest in these techniques, their application to biomedical field is still somewhat limited to straightforward application of sparse representation classification, or learning of multiple separate dictionaries, one for each class followed by representation residual-based loss functions. Therefore there is still motivation for the design of methods that leverage the capabilities of dictionary learning and sparse coding using joint discriminative-generative approaches.

In this work we introduce discriminative dictionary techniques that utilize class label consistency into spatially localized ensemble sparse analysis classification that we denote as LC-SLESA. We utilize this method for breast cancer diagnosis in mammograms. Our premise is that spatially localized dictionaries that have been optimized using label consistency constraints \cite{jiang2013label}, will improve the classification accuracy of spatially localized sparse analysis.  

\subsection{Sparse Analysis}
Conventional sparse representation of signals has been a research area of considerable interest in recent years. In image classification tasks, the representation of an image is used to determine the class of the test image. Sparse analysis seeks to optimize an signal representation objective function with signal sparsity constraints. This function  contains a term that measures the difference between the image reconstruction and the test image. This difference is known as the reconstruction error or residual. A sparsity term is also included in the objective function to measure the sparsity of the computed solution. The reconstruction error term may be set to measure the test image exactly or a constraint may be defined to provide an upper bound for the reconstruction error. 
The success of sparse representation based classification is due to the fact that a high-dimensional image can be represented or coded by few representative samples. Sparse representation based classification has two phases: coding and classification. Within the coding phase an image/signal is collaboratively coded over a dictionary of atom with some sparsity constraint.
In the classification performed based on the coding coefficients and the dictionary.

The dictionary could be predefined. For example, a dictionary may be made up of all training samples from all classes. This type of dictionary may not be able to represent test samples well due to the uncertain and noisy information in the original training images. Large number of dictionary atoms increase the coding complexity.

The optimization problem that sparse analysis seeks to solve is as follows:
Given samples from space $\mathbb{R}^d$, a dictionary $D \in \mathbb{R}^{d \times n}$ of signals partitioned by class, and a test signal $y \in \mathbb{R}^d$, sparse coding seeks to find a coding vector $x \in \mathbb{R}^n$. The test signal $y$ will be represented as a linear combination of the dictionary atoms with respect to the solution vector 
\begin{equation}
y = D x
\end{equation}

This is an ill-posed inverse problem. The problem is regularized by assuming sparse solutions; the sparsity of $x$ ultimately ensures that recovery is possible. Sparsity is represented by the $l_0$ norm and can be approximated by the $l_1$ norm, or $l_p$ norms with $p \in (0,1)$.
Sparse coding seeks to find a solution $\hat{x}$ by solving the following problem,
\begin{equation}
\hat{x}=\arg\min_x||\hat{x}||_0 \quad \text{subject to} \quad y = Dx
\end{equation}
Assuming a noisy signal, we can introduce the approximation tolerance $\epsilon$ and solve the following problem,
\begin{equation}
\hat{x}=\arg\min_x||\hat{x}||_0 \quad \text{subject to} \quad ||y - Dx|| < \epsilon 
\end{equation}

The objective function in equations (2) or (3) is optimized through sparse coding. Conventional sparse representation methods such as SRC \cite{Wright2009}, optimize an objective function of two terms, and use the original training images as the atoms in $D$. More recent works emphasize on dictionary design and task-specific optimization that we discuss next.

\subsection{Dictionary Learning}
Sparse representation seeks to represent a signal by linear combinations of dictionary atoms. While the objective is to achieve sparse solutions, the dictionary is a key factor that drives the optimization problem. The design of dictionary from training data has gained significant interest in recent years \cite{shrivastava2015generalized,yang2014sparse}. This area, known as dictionary learning seeks to construct the best dictionary possible to provide more efficient representation of classes for a specific task. Learning the dictionary has shown to lead to better representations versus using an unlearned dictionary \cite{yang2014sparse}. Dictionary learning can be categorized into three major areas \cite{tosic2011dictionary}: (i) probabilistic learning methods, (ii) clustering-based learning methods, and (iii) construction methods. The signal that will be represented, denoted by $y$, is of size $1 \times d$, the solution $\hat{x}$ is of size $n \times 1$ where $n$ is the number of atoms within the dictionary, and the dictionary $D$ is of size $d \times n$.   

The dictionary (sometimes referred to as the 'resource' database) consists of signals, also called atoms. Each atom makes up a column of the dictionary. The type, size, and design of the dictionary are important features. A dictionary may be pre-selected or trained (learned). The size of the dictionary has shown to have much effect on the representation formed. If the number of atoms within the dictionary ($n$) exceeds the dimension of the test signal ($d$), that is if $d < n$, the dictionary is said to be overcomplete. Studies show that an overcomplete/redundant dictionary leads to more sparse representation of signals \cite{Papyan2017}. 

\section{Label Consistent Spatially Localized Ensemble Sparse Analysis (LC-SLESA)} 
Based on our previous work on spatially localized ensemble sparse analysis (SLESA), here we introduce label consistency into the localized dictionaries and we denote this method by LC-SLESA. Our SLESA approach reduces the dimension of the feature vector, and integrate Bayesian decision learners to correct the estimation bias. At the same time, the ensemble classification we designed builds an integrated model by calculating the sparse representation of a block structure, thereby determining the lesion category. LC-SLESA aims to further improve the performance of SLESA by finding task-specific dictionaries that are consistent with the class labels of the training data.

\subsection{Spatially Localized Block Decomposition}
We divide each image into blocks  with size $m \times n$, represented as $I = [B^1, B^2, ... ,B^{NBL}]$, where $B^j$ denotes a block of each training image, and $NBL$ is the number blocks of an image. The dictionary $D^j$, where $j = 1,2,...,NBL$, is corresponding to the same position of the block $B^j$ for all $s$ images:
\begin{equation}
	D^j = [B_{1}^{j}, B_{2}^{j}, \cdots, B_{s}^{j}].
	\end{equation}
	
\subsection{Block-based Label Consistent KSVD for Dictionary Learning}
After the decomposition into spatially localized blocks, we learn $NBL$ discriminative dictionaries using the label consistent KSVD algorithm (denoted by LC-KSVD) proposed in \cite{jiang2013label}. This method adds task-specific constraints to the sparse representation problem to compute a single discriminative dictionary. 

In this work, we employ the LC-KSVD method to learn spatially localized dictionaries $D^j$.
We employed two variations of the objective function; LC-KSVD1 and LC-KSVD2. LC-KSVD1 adds a label consistent regularization term for the objective function and solves the following problem 
\begin{equation}
\arg \min_{D^j,A^j,x^j} ||Y^j - D^j X^j||_2^2 + ||Q^j - A^j X^j||_2^2 \quad s.t. \quad ||x_m^j||_0 \leq T, \quad \forall m \in [1, n].
\end{equation}
The added term is the discriminative sparse code error that forces signals from the same class to have similar sparse codes. $Q^j$ denotes the discriminative sparse codes for $Y^j$, $A^j$ is a linear transformation matrix, and $T$ is the sparsity threshold.

Furthermore, LC-KSVD2 adds a joint classification error and label consistent regularization term to the objective function. LC-KSVD2 is defined as
\begin{equation}
\arg \min_{D^j,A^j,W^j,x^j} ||Y^j - D^j X^j||_2^2 + ||Q^j - A^j X^j||_2^2 + ||H^j - W^j X^j||_2^2 \quad s.t. \quad ||x_m^j||_0 \leq T.
\end{equation}
The new term represents the classification error, $W^j$ denotes linear classifier parameters, and $H^j$ contains the class labels of the training data $Y^j$ \cite{jiang2013label}.

\subsection{Ensemble Classification}
At this stage we combine the individual spatially localized decisions to classify the test samples. For each test sample $y^j$ in $j$th block, we find the solution $x^j$ of the regularized noisy $l_1$-minimization problem using the dictionaries $\widehat{D^j}$ learned in the previous stage:
\begin{equation}
	\widehat{x}^j = \arg \min {||x^j||_1}\,\, \mbox{subject to}\,\, \Vert \widehat{D^j} x - y^j \Vert_2 \leq \epsilon 
	\label{eqt_block_src}
\end{equation}

We propose ensemble learning techniques in a Bayesian probabilistic setting as weighted sums of classifier predictions. We propose a decision function that applies majority voting to individual hypotheses (BBMAP) and an ensemble of log-likelihood scores computed from relative sparsity scores (BBLL).
\paragraph{Maximum a Posteriori decision function (BBMAP)}
The class label of each test image determined by the overall vote of the $NBL$ blocks-based classifiers. The predicted class label $\widehat{\omega}$ is given by
\begin{equation}
	\widehat{\omega}_{BBMAP} = \mathcal{F}_{BBMAP}(\widehat{x}) \doteq \arg{\max_i{pr(\omega_i|\widehat{x})}},
\label{eqt_BBMAP}
\end{equation} 
The probability for classifying $\widehat{x}$ into class  $\omega_i$ is 
\begin{align}
pr(\omega_i|\widehat{x}) &= \sum_j^{NB} ND_{\omega^j_i} / NB \\
ND_{\omega^j_i} &=
\begin{cases}
 1, & \text{if } \widehat{x}^j \in i\text{th class} \\
 0, & \text{otherwise}
			   \end{cases},
\end{align}
where $ND_{\omega^j_i}$ is an indicator function, and its value is determined by the decision of each classifier.
\paragraph{Log likelihood sparsity-based decision function (BBLL-S)}
A likelihood score based on the relative sparsity scores $\Vert \delta_m ( \widehat{x}^j ) \Vert_1$, $\Vert \delta_n ( \widehat{x}^j ) \Vert_1$ calculated at the sparse representation stage of each classifier
\begin{equation}
LLS(\widehat{x}) = -\log{\frac{\Vert \delta_m ( \widehat{x}^j ) \Vert_1}{\Vert \delta_n ( \widehat{x}^j ) \Vert_1}}\,\,\,	\begin{cases}
      										     \geq 0, & \widehat{x}^j \in m\text{th class} \\
     										     < 0, & \widehat{x}^j \in n\text{th class}
    										    \end{cases}.
\label{eqt_BBLL}
\end{equation}
The expectation of $LLS(\widehat{x})$ for all classifiers that we denote by $ELLS$ is estimated over the individual classification scores obtained by \eqref{eqt_BBLL}
\begin{align}
ELLS \doteq E\{LLS(\widehat{x})\} &= \frac{1}{NB} \sum_j^{NB} LLS(\widehat{x}^j)  \nonumber \\ &= -\frac{1}{NB}  \left[ \sum_j^{NB} \log{\Vert \delta_m ( \widehat{x}^j ) \Vert_1} - \sum_j^{NB} \log{\Vert \delta_n ( \widehat{x}^j ) \Vert_1} \right],
\end{align} 
We apply a sigmoid function $\varsigma(.)$ to determine the state of $\widehat{\omega}$ through the decision threshold $\tau_{LLS}$.
\begin{equation}
\widehat{\omega}_{BBLL} = \mathcal{F}_{BBLL}(\widehat{x}) \doteq \varsigma(ELLS(\widehat{x})-\tau_{LLS}).
\end{equation}

\section{Experiments and Discussion}
In this section we describe our experiments and report results produced by our approach and by widely used convolutional neural networks \cite{krizhevsky2012imagenet,Szegedy2015} to accommodate comparisons.

\subsection{Data}
We evaluated our CAD techniques for separation of breast lesions into two classes: malignant and benign. The training and testing data were obtained from the Mammographic Image Analysis Society (MIAS) database that is available online \cite{Oliver2010}. The resolution of the mammograms is 200 micron pixel edge that corresponds to about 264.58 $\mu m$ pixel size, and the size of each image is $1024 \times 1024$px after clipping/padding. MIAS contains 322 MLO scans from 161 subjects. Our goal is to characterize the lesion type, therefore we utilized 66 benign and 51 malignant mammograms for performance evaluation. 

ROI selection is applied first, in order to prepare the data for block decomposition. We need to ensure that the majority of the blocks cover the lesion to improve the accuracy. Hence, we designed our system so that the lesion ROI sizes are greater than or equal to the analysis ROI size. Our method reads-in the centroid and radius of each mass from the provided radiological readings. It uses these two values to automatically determine a minimum bounding square ROI and to select the masses that satisfy the size criterion. We used a ROI size criterion of $64 \times 64$px, resulting in 36 benign and 37 malignant lesions. These ROIs contain sufficient visual information, while preserving a big part of the data samples. We performed 10-, 20- and 30-fold cross-validation to study the effect of the size of folds on performance.

\subsection{Convolutional Neural Networks}
In this part of our experiments we implemented CNN classification using the Alexnet \cite{krizhevsky2012imagenet} and Googlenet \cite{Szegedy2015} architectures with transfer learning. Both networks were pre-trained on Imagenet that is a database of 1.2 million natural images.

We applied transfer learning to each network in slightly different ways.
To adjust Alexnet to our data, we replaced the pre-trained fully connected layers with three new fully connected layers. We set the learning rates of the pre-trained layers to 0 to keep the network weights fixed, and we trained the new fully connected layers only. In the case of Googlenet, we set the learning rates of the bottom 10 layers to 0, we replaced the top fully connected layer with a new fully connected layer, and we assigned a greater learning rate factor for the new layer than the pre-trained layers.

To provide the networks with additional training examples, we employed data resampling using randomly-centered patches, followed by data augmentation by rotation, scaling, horizontal flipping, and vertical flipping.
Finally, we applied hyperparameter tuning using Bayesian optimization to find the optimal learning rate, mini-batch size and number of epochs.

We first applied 10-, 20-, and 30-fold cross-validation to $64\times64$px ROIs. Because deep networks can learn information from the edges of lesions and not just the texture, we also decided to test our method on $256\times256$px ROIs of all lesions (66 benign and 51 malignant) to improve the classification performance. We report the results of our cross-validation experiments in Table \ref{tab_MIAS_CNN}. 
We note that Alexnet produces the top ACC of 67.65\% and the top AUC of 63.04\% for 30-fold cross-validation and for all $256 \times 256$px ROIs.

\begin{table}[H]
\caption{
Classification performance for breast lesion characterization using convolutional  neural network classifiers} \label{tab_MIAS_CNN}
\begin{center}
\resizebox{\textwidth}{!}{
\begin{tabular}{| p{55pt} | c | c | c | c | c | c |}
\hline
Method & k-Fold CV & ROI Size & TPR $(\%)$  & TNR $(\%)$  & ACC $(\%)$  & AUC $(\%)$	\\\hline
\multirow{6}{*}{Alexnet} & 10 & $64 \times 64$ &  50.0  & 58.33 & 54.17 & 47.69 \\\cline{3-7} & & $256 \times 256$ &  56.86  & 72.55 & \textbf{64.71} & \textbf{62.19} \\
\cline{2-7} & 20 & $64 \times 64$ & 44.44 & 69.44 & 56.94 & 52.55 \\
\cline{3-7} & & $256 \times 256$  &  47.06  & 84.31 & \textbf{65.69} & \textbf{60.7} \\
\cline{2-7}
& 30 & $64 \times 64$ & 38.89 & 72.22 & 55.56 & 53.97 \\
\cline{3-7} & & $256 \times 256$  & 58.82 & 64.71 & 61.77 & 60.29 \\ \hline
\multirow{6}{*}{Googlenet} & 10 & $64 \times 64$ &  25.0  & 83.33 & 54.17 & 47.21 \\\cline{3-7} & & $256 \times 256$  & 64.71 & 58.82 & 61.77 & 57.86 \\
\cline{2-7} & 20 & $64 \times 64$ & 58.33 & 50.0 & 54.17 & 50.22 \\
\cline{3-7} & & $256 \times 256$  &  62.75  & 62.75 & 62.75 & 61.5 \\
\cline{2-7} & 30 & $64 \times 64$ & 58.33 & 50.00 & 54.17 & 50.96 \\
\cline{3-7} & & $256 \times 256$  & 66.67 & 68.63 & \textbf{67.65} & \textbf{63.04} \\
\hline 
    \end{tabular}
}  \end{center} 
\end{table}

\subsection{LC-SLESA}
Next, we validated our block-based ensemble classification system.  We performed 10-, 20- and 30-fold cross-validation on these samples. We present results on characterization of lesions with minimum ROI size of $64 \times 64$ pixels that forms a dataset of 36 benign and 37 malignant lesions.

\begin{figure}[h]
\centering
\begin{tabular}{c c}
\includegraphics[width=.4\textwidth,clip,trim=0in 0in 0in 0in]{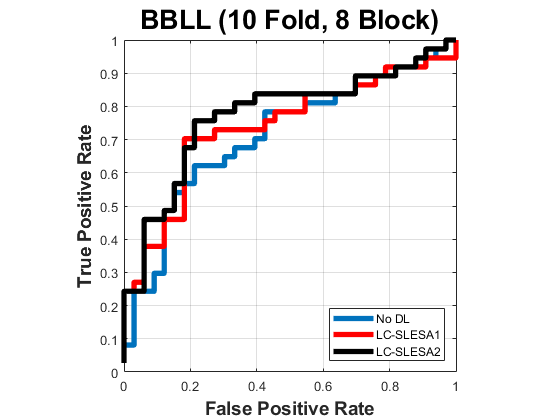} &
\includegraphics[width=.4\textwidth,clip,trim=0in 0in 0in 0in]{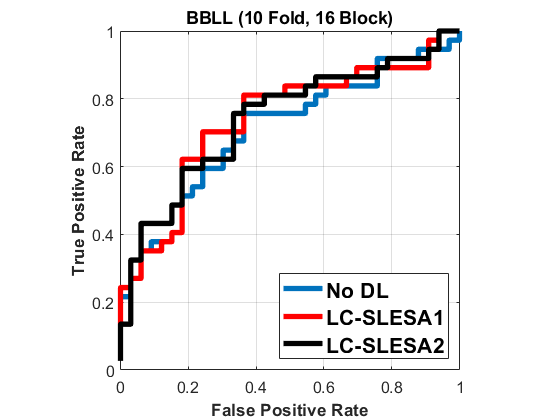} \\
\includegraphics[width=.4\textwidth,clip,trim=0in 0in 0in 0in]{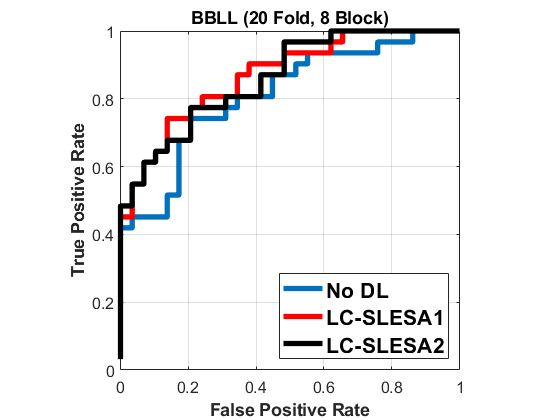} &
\includegraphics[width=.4\textwidth,clip,trim=0in 0in 0in 0in]{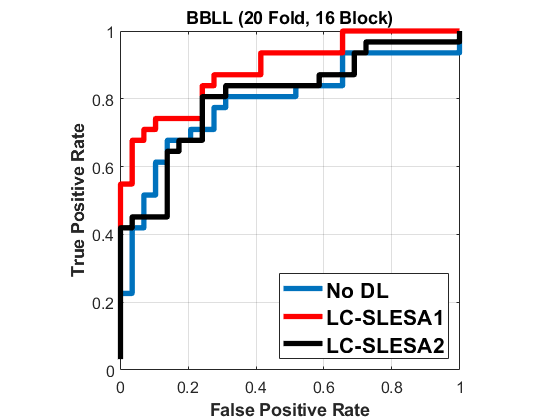} \\
\includegraphics[width=.4\textwidth,clip,trim=0in 0in 0in 0in]{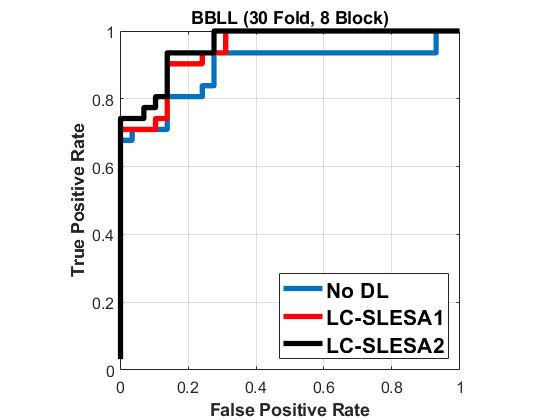} &
\includegraphics[width=.4\textwidth,clip,trim=0in 0in 0in 0in]{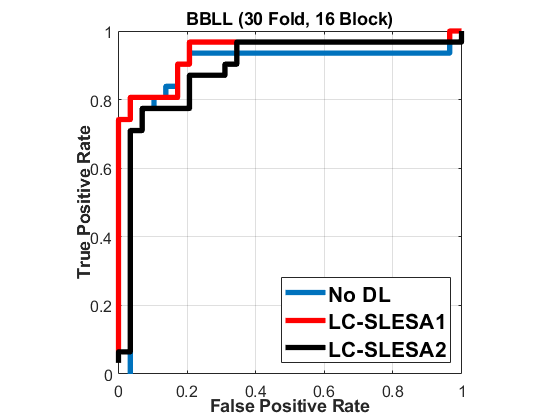} \\
\end{tabular}
\caption{ROC plots for $8 \times 8$ and $16 \times 16$ block sizes using the proposed block-based ensemble method with BBLL decision functions and 10-fold (top row), 20-fold (second row) and 30-fold (bottom row) cross-validation.}
\label{fig_MIAS_64_ROC_BB_all_folds_8_16_block} 
\end{figure}

\begin{figure}[h]
   \centering
     \subfloat[No DL\label{subfig-1:No_DL_X1}]{%
       \includegraphics[width=0.49\textwidth,clip,trim=.75in .5in .75in .25in]{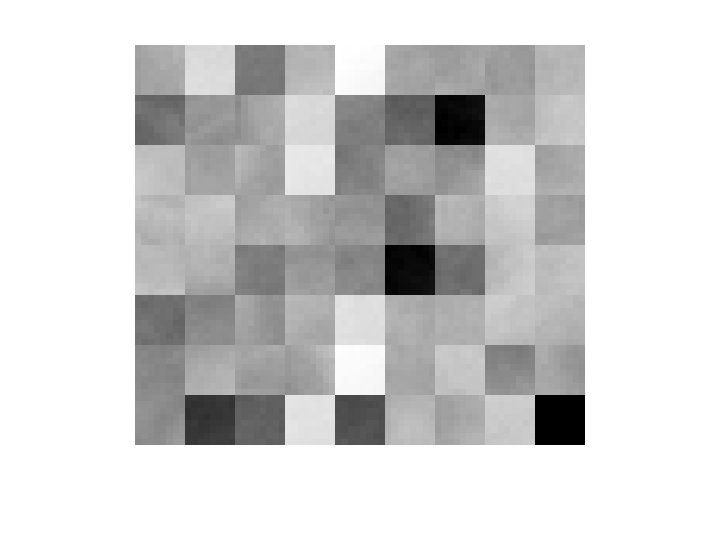}
     }
     \subfloat[LC-KSVD1\label{subfig-2:LCKSVD1_Dictionary}]{%
       \includegraphics[width=0.49\textwidth,clip,trim=.75in .5in .75in .25in]{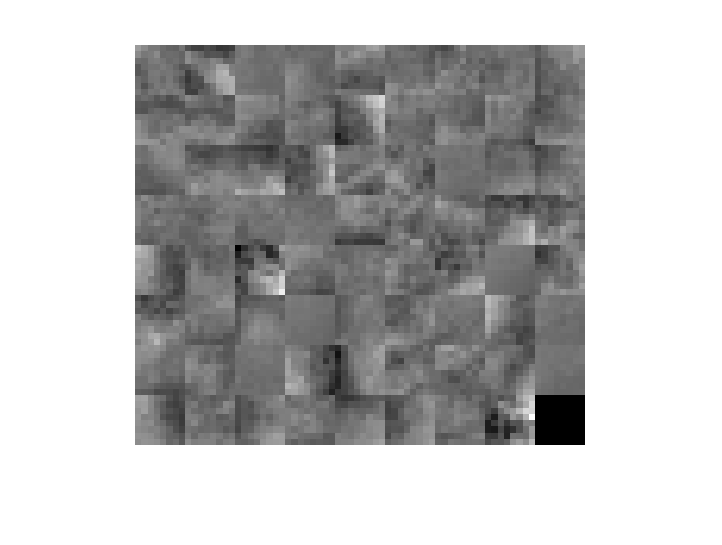}
     }\\
     \subfloat[LC-KSVD2\label{subfig-2:LCKSVD2Dictionary}]{%
       \includegraphics[width=0.49\textwidth,clip,trim=.75in .5in .75in .25in]{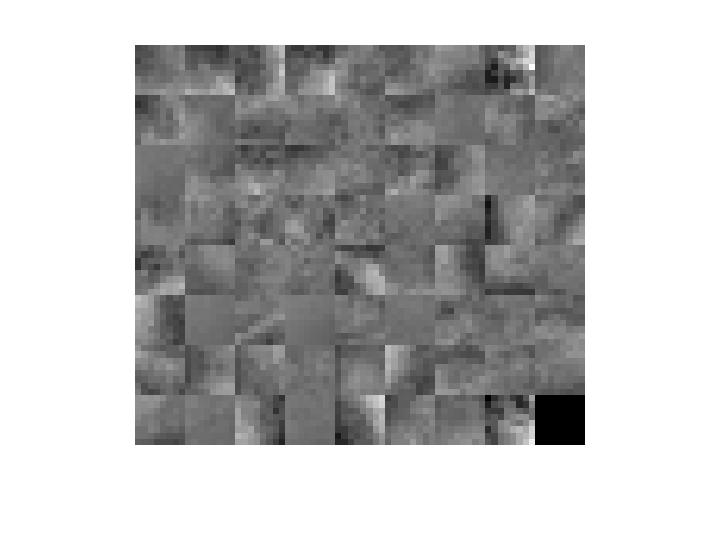}
     }
     \caption{Dictionary comparison example for (a) SLESA without dictionary learning, (b) LC-SLESA1, and (c) LC-SLESA2 methods.}
     \label{fig:Dictionaries}
   \end{figure}

Table \ref{tab_MIAS_BB_LCKSVD_rand_state} contains the classification rates produced by our cross-validation experiments for multiple block sizes. In the first row of this table, the results were obtained from a single block that is equivalent to conventional SRC analysis \cite{Wright2009}. 
We note that the accuracy increases when the number of folds increases for the same ROI size. The highest accuracy by using 10-fold cross-validation is 75.71\% for $32 \times 32$ block size for LC-SLESA1 with BBLL decision function. The largest area under the curve for 10-fold CV is 77.31\% for $8 \times 8$ block size for LC-SLESA2 with BBLL decision function. For 20-fold cross-validation, the best accuracy is 80.00\% and AUC is 88.88\% for $8 \times 8$ block size for LC-SLESA1 with BBLL decision function. The best overall performance is obtained for 30-fold cross validation. The highest accuracy is 88.33\% for $16 \times 16$ block size for LC-SLESA1 with BBLL decision function, and the largest area under the curve is 95.88\% for $8 \times 8$ block size for LC-SLESA2 with BBLL decision function. There are 2 or 3 test samples in each fold when $k=30$. We also display the receiver operating curves by variations of the SLESA method in Figure \ref{fig_MIAS_64_ROC_BB_all_folds_8_16_block} for $8 \times 8$ and $16 \times 16$ block lengths. These graphs lead to the same observations that we made from Table \ref{tab_MIAS_BB_LCKSVD_rand_state}. In addition, Figure \ref{fig:Dictionaries} displays an example of dictionaries based on the original training images, learned by LS-KSVD1 and learned by LC-KSVD2. We see that the label consistent algorithms learn structural features of the lesions.

\begin{table}[h]
\caption{Classification performance for breast lesion characterization using ensembles of block-based sparse classifiers with dictionary learning (ROI size: $64 \times 64$)}
\label{tab_MIAS_BB_LCKSVD_rand_state}
\begin{center}
\resizebox{\textwidth}{!}{
\begin{tabular}{| p{55pt} | c | c || c | c || c | c || c | c |}
\hline
Method & k-Fold & Block Size & SLESA & SLESA & LC-SLESA1 & LC-SLESA1  & LC-SLESA2 & LC-SLESA2 \\			  
& CV & & ACC $\%$& AUC $\%$ & ACC $(\%)$  & AUC $(\%)$  & ACC $(\%)$ & AUC $(\%)$ \\  \hline
BBMAP-S & 10 & $64 \times 64$& 57.14&54.55 &47.14 &53.51 &51.43 &47.50\\\cline{3-9}
& & $32 \times 32$  & 64.29 &64.29 &67.14 & 68.14&55.71 &54.63 \\\cline{3-9}
& & 	$16 \times 16$ & 70.00 &69.70 & 70.00&69.70 &70.00 &67.70 \\\cline{3-9}
& & $8 \times 8$ & 64.29&63.64 &70.00 &69.70 &70.00 &69.70 \\\cline{3-9}
\hline
BBLL-S & 10 & $64 \times 64$ & 60.00& 62.33& 55.71& 42.51&57.14 &52.74\\\cline{3-9}
& & $32 \times 32$  & 64.29&60.44 &\textbf{75.71} &73.96 & 62.86&57.66\\\cline{3-9}  
& & $16 \times 16$  & 68.57 &70.84 &72.86 &73.96 &70.00 &73.96\\\cline{3-9}  
& & $8 \times 8$   & 70.00& 71.42& 71.40& 73.87& 70.00& \textbf{77.31}\\\cline{3-9}
\hline  \hline
BBMAP-S & 20 & $64 \times 64$& 46.67& 42.38&45.00 &40.16 &60.00 &55.73 \\\cline{3-9}
& & $32 \times 32$ & 58.33& 54.84& 53.33& 50.17& 55.00&52.84\\\cline{3-9}  
& & $16 \times 16$& 75.00& 76.64& 76.67& 76.42& 71.67&71.08\\\cline{3-9}  
& & $8 \times 8$   & 75.00&76.64 &70.00 &69.19 &75.00 &75.19\\\cline{3-9}
\hline
BBLL-S & 20 & $64 \times 64$ & 63.33&53.73 &53.33 &48.05 &61.67 &58.18	\\\cline{3-9}
& & $32 \times 32$   & 60.00& 55.95& 55.00& 54.62& 68.33&61.07\\\cline{3-9}
& & 	$16 \times 16$ & 76.67& 78.87&\textbf{80.00} &\textbf{88.88} &76.67 &80.65\\\cline{3-9}
& & $8 \times 8$ &  76.66& 81.20&78.33 &86.99 &78.33 &86.21\\\cline{3-9}
\hline \hline
BBMAP-S & 30 & $64 \times 64$ & 51.67& 46.50& 46.67&40.71 &56.67 &50.61\\\cline{3-9}
& & $32 \times 32$ &  53.33& 48.83& 60.00& 55.51& 50.00&46.38\\\cline{3-9}
& & 	$16 \times 16$ & 85.00& 85.65&85.00 &85.65 &85.00 &85.65 \\\cline{3-9}
& & $8 \times 8$ & 83.33& 82.09&70.00 &69.19 &85.00 &85.65\\\cline{3-9}
\hline
BBLL-S & 30 & $64 \times 64$ &65.00& 58.62& 53.33& 43.83&60.00 &61.18 \\\cline{3-9}
& & $32 \times 32$ &  68.33& 65.41&66.67 &60.07 &66.67 &62.63\\\cline{3-9}  
& & $16 \times 16$ & 86.67& 88.21& \textbf{88.33}&93.66&83.33 &88.88\\\cline{3-9}  
& & $8 \times 8$   & 83.33&89.10 &85.00 &94.66&85.00 &\textbf{95.88}\\\cline{3-9}
 \hline
    \end{tabular}
}  \end{center}
\end{table}

\section{Conclusion}

We proposed discriminative localized sparse representations for classifying breast masses into benign and malignant states. This approach introduces discriminative capability into the generative method of sparse representation for classification. As we observed in our experiments, this approach improves the classification accuracy of integrative sparse analysis and accomplishes an area under the ROC of about 95.88\% for 30-fold cross-validation.

\section*{Acknowledgments}
The authors acknowledge the support by the National Institute of General Medical Sciences of the National Institutes of Health under Award Number SC3GM113754. They also acknowledge the support by Delaware CTR-ACCEL (NIH U54GM104941) and the State of Delaware.

{\small
\bibliographystyle{splncs04}
\bibliography{Dictionary_Learning_Sparse_Analysis_Mammography_ISMCO_2020}
}
\end{document}